\documentclass[11pt,amsmath,floats,floatfix,nofootinbib,superscriptaddress]{revtex4}
\usepackage{graphicx} \usepackage{bm} \usepackage{epsfig}
\usepackage{pstricks}
\usepackage{setspace}
\usepackage{graphicx}
\usepackage{bm}
\usepackage{epsfig}
\usepackage{color}
\usepackage{colordvi}
\newcommand{\beq}{\begin{equation}}
\newcommand{\eeq}{\end{equation}}
\newcommand{\bea}{\begin{eqnarray}}
\newcommand{\eea}{\end{eqnarray}}
\newcommand{\bit}{\begin{itemize}}
\newcommand{\eit}{\end{itemize}}

\newcommand{\nn}{\nonumber}

\begin{document}

\title{Next to leading order spin-orbit effects in the motion of inspiralling compact binaries}
\author{Rafael A. Porto}
\affiliation{Kavli Institute for Theoretical Physics, University of California, Santa Barbara CA 93106, USA}

\begin{abstract}
Using effective field theory (EFT) techniques we calculate the next-to-leading order (NLO) spin-orbit contributions to the gravitational potential of inspiralling compact binaries. We use the covariant spin supplementarity condition (SSC), and explicitly prove the equivalence with previous results by Faye et al. in \url{gr-qc/0605139}. We also show that the direct application of the Newton-Wigner SSC at the level of the action leads to the correct dynamics using a canonical (Dirac) algebra.
This paper then completes the calculation of the necessary spin dynamics within the EFT formalism that will be used in a separate paper to compute the spin contributions to the energy flux and phase evolution to NLO.
\end{abstract}
\date{\today}

\maketitle
\newpage
\tableofcontents

\newpage

\section{Introduction}

LIGO \cite{ligo}, VIRGO \cite{virgo} and eventually LISA \cite{lisa} and Einstein Telescope \cite{etele} expect to detect radiation from inspiralling binary systems, and building templates for these events has become increasingly important \cite{gwaves1,gwaves2}. While for late stages of the inspirals numerical techniques are needed \cite{luis,frans,carlos}, for the early stages we may rely on the Post-Newtonian (PN) approximation as an expansion in small velocities $(v/c)\ll 1$ (see \cite{review} for a review and further references.)

As argued in the literature  \cite{pan, buo3, hughes,arun}, the expectation that black holes may be close to maximally rotating in the binaries \cite{McClintock} produces a significant impact in the accuracy of gravitational wave templates. Therefore, if not for detection, parameter extraction justifies the need for physical templates that include spin contributions at higher PN orders beyond the leading effects. In addition, PN corrections including spin are relevant for comparison between analytic results and numerical simulations \cite{lousto}.\\

 In order to produce such templates an important building block is the gravitational potential, $V$, responsible for the dynamics of the bodies in the binary system. The leading order (LO) spin-orbit and spin-spin contributions to $V$ have been known in the literature for quite some time \cite{wald, barker,dass, ehlers, kidder1, kidder2,poisson}, and spin precession has already been observed in some binary pulsars, e.g. \cite{prec}. On the other hand, only recently the next-to-leading order (NLO) spin-orbit \cite{owen, bbf, damour}, spin(1)spin(2) \cite{eih,proc, comment,nrgrss, Schafer3pn} and spin(1)spin(1) \cite{nrgrs2,Schafer3pn2,Hergt} contributions have been computed, albeit with rather dissimilar techniques.
 
The NLO spin-orbit effects (2.5PN) were computed in \cite{owen, bbf} within the `standard' PN approach \cite{review}, namely by obtaining the metric using Bailey-Israel-Dixon's spin-dependent stress energy tensor \cite{dixon} and then solving Mathisson-Papapetrou equations \cite{papa}. On the other hand, in \cite{damour} the frequency of spin precession was directly obtained in ADM coordinates by a `suitable redefined constant-magnitude spin vector' and subsequently a Hamiltonian was derived that was shown to be equivalent to the results in \cite{bbf}. This is a very useful trick that we borrow in this paper, although we will obtain the frequency of precession using a completely different approach.\\
    
The NLO spin(1)spin(2) effects (3PN) were first obtained in \cite{eih,proc,comment} using a new formalism based on the application of Effective Field Theory (EFT) techniques introduced in \cite{nrgr}, where it was coined Non-Relativistic General Relativity (NRGR), and extended to include spinning constituents in \cite{nrgrs}. An EFT framework\footnote{See \cite{iraeft} for a review on EFT techniques.} has shown to be extremely powerful in many realms of {\it classical} gravitational (and non-gravitational) physics. For example, it has been used to compute the NNLO spin-independent contribution to the potential (2PN) \cite{andi1}; absorption effects for compact objects in binary systems and time dependent backgrounds \cite{dis1,dis2}; the electromagnetic \cite{adam} and gravitational self-force on extended objects \cite{chad1}; radiation-reaction effects in the extreme mass ratio limit \cite{chad2}; corrections to thermodynamic quantities for caged black holes \cite{cbh,andi2}; deviations from General Relativity at the nonlinear level \cite{mag}; and recently applied to the study of cosmological perturbations \cite{eftfluid}. 

The radiation sector of NRGR has been developed further in \cite{rad1, spinrad}. The purpose of this paper is thus to compute the NLO spin-orbit corrections to the gravitational potential\footnote{See \cite{delphine} for an alternative derivation of the NLO spin-orbit potential.} (within our set of conventions), which together with the spin-spin results in \cite{eih,comment,proc,nrgrss,nrgrs2} and source multipole moments computed in \cite{rad1, spinrad} allow us to obtain all the spin contributions in the energy flux and phase evolution to 3PN order. This will be reported in a separate paper\footnote{A comparison with the recent results for the NLO spin-orbit contributions to the radiated power in \cite{buo2} will be performed in \cite{spinrad2}.} \cite{spinrad2}.\\

A crucial aspect of the calculations dealing with spin in General Relativity is the choice of spin supplementarity conditions (SSCs) \cite{pryce,tulc} and, in a Lagrangian/Hamiltonian framework, the resulting (Dirac) algebra in the reduced phase space. This algebra is expected to be quite cumbersome in generic SSCs, already in a Minkowski background \cite{regge}. Nevertheless, the results in \cite{eih,comment} were obtained using the Newton-Wigner (NW) SSC at the level of the action, which was shown to be a correct procedure in \cite{proc} using standard EFT power-counting techniques. As we shall show in this paper, the NW SSC also leads to a canonical structure at linear order in spin, at least to NLO. This is consistent with the results in \cite{brb} (although formally the comparison only applies in the test-particle limit.) 

Subsequently the potential in \cite{eih,comment} was re-derived using a Routhian approach, introduced in \cite{proc}, where the (covariant) SSC is only imposed at the late stages of the calculation so that we may never depart from a canonical algebra, and full consistency was found \cite{nrgrss}. The NLO spin(1)spin(2) Hamiltonian has also been computed in \cite{Schafer3pn} in ADM coordinates, and the equivalence with the results in \cite{eih} was shown in \cite{comment}.\\

The basic idea of our approach is the systematic separation of the relevant scales $(\lambda_{\rm fs} \ll \lambda_{\rm p} \ll \lambda_{\rm r})$ in the two body problem; from finite size effects ($\lambda_{\rm fs} \sim Gm$ for compact objects), to the potential ($\lambda_{\rm p} \sim r$) and gravitational radiation ($\lambda_{\rm r} \sim \frac{r}{v}$) scales (for a review see \cite{nrgrLH}.) For the theory of potentials the NRGR prescription consists on computing all possible (Feynman) diagrams (without external radiation gravitons) scaling with a definite power of $v$. For spinning extended objects, the `rules' are derived from an action (Routhian) principle and are listed in \cite{nrgr,nrgrs,nrgrss,nrgrs2}. The EFT formalism is therefore significantly more efficient than the (more traditional) PN methods since it maps complex integrals into the computation of Feynman diagrams, uses textbook regularization techniques, and it is especially suited to handle spin degrees of freedom. Feynman diagrams allow for a very natural systematization and (physical) visualization of the computation, and moreover it can be automatized almost entirely using {\it Mathematica} code \cite{nrgrwork}. As a result, the calculation of the NLO spin(1)spin(2) potential was no more involved than obtaining the Einstein-Infeld-Hoffmann Lagrangian as in \cite{nrgr}, contrary to the intricate approach of \cite{Schafer3pn}.\\

The NLO spin(1)spin(1) potential (also at 3PN) was computed in \cite{nrgrs2}. Most of the spin(1)spin(1) effects are due to (self-induced) finite size contributions \cite{nrgrs} (and an extra term responsible for the preservation of the SSC \cite{proc,nrgrs2}.) In the spirit of EFTs these terms are encoded in higher dimensional operators in the effective action (constrained only by the symmetries of the theory) up to unknown coefficients that are `matched' with a given observable in the complete theory \cite{nrgr}, for example the quadrupole moment of a spinning black hole \cite{nrgrs,nrgrs2}. Once again the gravitational potential follows as a sum of Feynman diagrams of definite PN order which greatly simplifies the treatment of finite size corrections. These results were corroborated in \cite{Schafer3pn2, Hergt}.\\ 

The Hamiltonian of \cite{Schafer3pn2,Hergt} was partially obtained by imposing Poincare invariance, which is not sufficient to determine all the coefficients, and the extra terms were fixed `by a three-dimensional covariant ansatz for the source terms of the Hamilton constraint.' As argued in \cite{Schafer3pn2}: `the arguments that lead to a unique fixation of the coefficients are, however, far from being straightforward.'

On the other hand, the determination of the potential in \cite{nrgrs2} is straightforward once a (very natural) higher dimensional operator is included. Moreover, using EFT power counting techniques one can easily show that one (and only one) operator is sufficient to NLO \cite{nrgrs, nrgrs2}, whose overall (Wilson) coefficient is directly related to the spin quadrupole of the Kerr black hole.

Truth be told, the advantage of the Hamiltonian in \cite{Schafer3pn2} is that the spin tensor is fully reduced, and one only deals with a three-vector with canonical brackets. In our formalism we bypass the complications of the (Dirac) algebra in the covariant SSC by reducing to a three-vector as the final step in the calculation. Therefore, in the intermediate steps we have an extra three-vector degree of freedom, ${\bf S}^{(0)i} \equiv S^{0i}$, with non-vanishing brackets (see Eq. (\ref{als0})), that is only solved for via the SSC once the equations of motion are obtained (see Eqs. (\ref{alssc} - \ref{spind}).)
This however does not represent any major difficulty, no more than dealing with an antisymmetric spin tensor obeying a canonical algebra\footnote{One can also perform a transformation that reduces the spin 3-vector algebra to standard brackets \cite{Hergt}.}.\\

Ultimately, given the differences in approaches, the agreement found in all cases gives us strong confidence the results are correct. In this paper we add yet another piece of evidence for the validity of the spin potentials by computing the NLO spin-orbit effects using the EFT framework of NRGR. The computation proceeds systematically as a sum of Feynman diagrams. Hence first we sketch\footnote{We encourage the reader to consult the NRGR literature for more details \cite{nrgr,nrgrs,eih,proc,nrgrLH,nrgrss,nrgrs2}.} the ingredients that determine the necessary Feynman rules before we embark in the computation of the diagrams, that we divide into terms linear and quadratic in $G$ (Newton's constant.) Divergences are handled by textbook regularization techniques that result in finite integrals. After the potential is obtained we perform a spin redefinition that takes the spin dynamics into precession form from which we  read off the precession frequency. The details are summarized in an appendix. 
Then it is straightforward to show that our result resembles those in \cite{bbf} since both frequencies, ours and the one given in \cite{bbf}, are related by a total time derivative. Following \cite{damour} one can then construct a Hamiltonian whose equations of motion turn out to be canonically related to those in \cite{bbf}, and the equivalence is thus proven.

In the last section we show that imposing the NW SSC at the level of the action also leads to the correct dynamics in a canonical manner. Remarkably, the transformation found in \cite{regge} between covariant and NW SSCs in a flat background applies to our case once re-written in a locally flat frame and with derivatives taken with respect to the `free falling' proper time. Therefore, here we extend the results in \cite{brb} to the case of self-gravitating objects, as long as we ignore spin(1)spin(1) (finite size) terms\footnote{Imposing the NW SSC does not lead to canonical brackets in the spin(1)spin(1) sector since it knocks off a necessary extra term in the Routhian, proportional to the SSC, which guarantees its preservation upon time evolution \cite{proc,nrgrss,nrgrs2}.}. 

\section{Spin effects in the Effective Field Theory approach}

To include spin effects in the EFT formalism the following action/Routhian was introduced\footnote{A similar expression was advocated in \cite{yee}.} \cite{nrgrs,eih,proc,nrgrss,nrgrs2}
\begin{equation} \label{actR}
{\cal R} =-\sum_{q=1,2} \left( m_q \sqrt{u^\alpha_{q}u_{q\alpha }} + 
\frac{1}{2}S_{q}^{ab}\omega_{ab\mu} u^\mu_{q} +\ldots\right)
\end{equation}
where the dots include ${\cal O} ({\bf S}^2)$ corrections that are not relevant for the present paper\footnote{In the approach of \cite{proc,nrgrss,nrgrs2} we were able to sweep most of the technicalities of the preservation of the SSC into an extra term in Eq. (\ref{actR}) proportional to spin(1)spin(1) that we can ignore at linear order in spin and also for spin(1)spin(2) effects as it was shown in \cite{eih,comment,nrgrss}.}.

The equations of motion derive from
\begin{equation}
\frac{\delta }{\delta x^\mu}\int {\cal R} d\lambda=0, \;\;\; \frac{d
S^{ab}}{d\lambda} = \{S^{ab},{\cal R}\}, \;\;\; \frac{d
S^{ab}}{d\lambda} = \{V, S^{ab}\}\label {eomV},
\end{equation}
where the potential is given by $V= -{\cal R}$ 
and 
\bea
\{S^{ab},S^{cd}\} &=& \eta^{ac} S^{bd}
+\eta^{bd}S^{ac}-\eta^{ad} S^{bc}-\eta^{bc}
S^{ad}. \label{als}
\end{eqnarray}

Following the NRGR prescription \cite{nrgr,nrgrs} we need to compute all possible diagrams linear in spin that enter at ${\cal O}(v^5)$ in the gravitational potential. We use reparamaterization invariance to select $\lambda=x^0=t$ such that $v^\mu \equiv \frac{dx^\mu}{dt}=(1,{\bf v})$. Then
each diagram contributes a term in the effective action/Routhian which is related to the potential via $i \int dt {\cal R}_{\rm diag} = -i \int dt V_{\rm diag}$ \cite{nrgr,nrgrs,proc,nrgrss,nrgrs2}. The necessary Feynman rules are \cite{nrgrs,eih,nrgrss,nrgrs2}
\bea
L_{\rm sg}^{v^2} &=& \frac{1}{2m_p}H_{i0,k}S^{ik},\label{sgnr1}\\
L^{v^3}_{\rm sg} &=& \frac{1}{2m_p}\left(H_{ij,k}S^{ik}{\bf v}^j + H_{00,k}S^{0k}\right),\label{sgnr15}\\
L^{v^4}_{\rm sg} &=& \frac{1}{2m_p}\left(H_{0j,k}S^{0k}{\bf v}^j +
H_{i0,0}S^{i0}\right)\\
L^{v^5}_{\rm sg} &=& \frac{1}{2m_p} H_{ik,0}S^{k0} {\bf v}^i
\label{sgnr2}, 
\eea
for the terms linear in the metric, whereas to ${\cal O}(H^2)$ 
\bea
L^{v^4}_{SH^2} &=& \frac{1}{4m^2_p}S^{ij} H^{\lambda}_j \left(H_{0\lambda,i} -
H_{0i,\lambda}\right) \label{sgh2v4}\\
L^{v^5}_{SH^2} &=& \frac{1}{4m^2_p}\left[S^{ij} H^l_j \left(H_{kl,i} -
H_{ki,l}\right){\bf v}^k + S^{i0}\left(H_{00}H_{00,i}+H^l_iH_{00,l}\right)\right]\label{sgh2v5},
\eea
where we only kept the terms that lead to non-zero contractions at the desired order. The field $H_{\mu\nu}$ represents the potential graviton responsible for the gravitational binding of the binary system \cite{nrgr}. 

To compute spin-orbit effects we also need the Feynman rules for the spinless couplings. These are \cite{nrgr}
\bea
L_{m}^{v^0} &=&- \frac{m}{2m_p}H_{00}\\
L^{v^1}_{m} &=& -\frac{m}{m_p} H_{0i}{\bf v}^i\\
L_{m}^{v^2} &=& -\frac{m}{2m_p}\left(H_{ij}{\bf v}^i{\bf v}^j + \frac{1}{2} H_{00} {\bf v}^2\right)\label{lv2m}\\
L_{m}^{v^3} &=& -\frac{m}{2m_p} H_{0i} {\bf v}^i {\bf v}^2 
\eea
and for the non-linearities
\bea
L_{mH^2}^{v^2} &=& \frac{m}{8m_p^2}H_{00}H_{00}\label{h2m}\\
L^{v^3}_{mH^2} &=& \frac{m}{2m_p^2} H_{00}H_{0i}{\bf v}^i\label{h2mv}.
\eea

In all these expressions $m_p \equiv \frac{1}{\sqrt{32\pi G}}$.  We encourage the reader to consult the literature for more detailed discussions
\cite{nrgr,nrgrs,eih,proc,nrgrLH, nrgrss,nrgrs2}.

\subsection{The (covariant) spin supplementarity condition}

In the Routhian formalism described above the (covariant) SSC is only enforced once the equations of motion are obtained using Eq. (\ref{eomV}). However, at NLO we need to account for corrections in the spin algebra due to gravitational effects in the transformation between local and PN frames. Since we work with the spin in the local frame, the covariant SSC has the form
\beq
S^{i0}_1 = S^{ij}_1 \frac{{\tilde{\bf v}}^j_1}{{\tilde v}^0_1}
\eeq
where
\bea
\label{tildv}
\tilde {\bf v}^{a=i}_1 &=& {\bf v}^i_1 \left(1+\frac{Gm_2}{r}\right) - \frac{2Gm_2}{r} {\bf v}^i_2, \\
\tilde v^{a=0}_1 &=& \left(1-\frac{Gm_2}{r}\right) \label{tild0}
\eea
is the particle's velocity in a locally flat frame, $v^a \equiv e^a_\mu v^\mu$, and
\bea
e^0_0({\bf x}_1) &=& 1 - \frac{G_N m_2}{r} + \cdots \\ 
e^k_0({\bf x}_1) &=& - 2 \frac{G_N m_2 {v}_2^k}{r} +\cdots  \\ 
e^i_j({\bf x}_1) &=& \delta^i_j \left(1+\frac{G_N m_2}{r}\right) +\cdots
\eea
with ${\bf r}={\bf x}_1-{\bf x}_2$. The spin algebra in Eq. (\ref{als}) then takes the form \cite{nrgrss}
\beq
\{{\bf S}^i,{\bf S}^j\} = -\epsilon^{ijk}{\bf S}^{k},~{\rm and}~~~\{S^{0i},{\bf S}^j\} = -\epsilon^{ijk}S^{0k}
\label{alssc}
\eeq
or equivalently\footnote{Notice that our conventions differ by an overall sign with those in \cite{Hergt}.} 
\beq
\label{als0}
\{{\bf S}^i,{\bf S}^{(0)j}\} = -\epsilon^{ijk}{\bf S}^{(0)k},
\eeq
where we introduced the spin (three-)vectors ${\bf S}^i = \epsilon^{ijk}S^{jk}$ and ${\bf S}^{(0)i} = S^{0i}$.\\

Even though we have to deal with ${\bf S}^{(0)}$ all the way to the end, it is straightforward to derive the equations of motion, for example for the spin dynamics, using Eq. (\ref{alssc}). If we write the ${\bf S}$ and ${\bf S}^{(0)}$ contributions to the gravitational potential as \beq \label{split} V_{{\bf S}^{(0)}} =  {\bf A}\cdot {\bf S}^{(0)},~{\rm and} ~~  V_{\bf S} = \omega_{\bf S}\cdot {\bf S}\eeq respectively, then using Eq. (\ref{eomV}) the spin equation of motion becomes
\beq
\label{spind}
\frac{d{\bf S}}{dt} = \omega_{\bf S}\times {\bf S} + \left(\frac{\tilde {\bf v}}{\tilde v^0}\times{\bf S}\right)\times {\bf A}
\eeq
in the covariant SSC, namely
\beq
\label{s0cov}
{\bf S}^{(0)} =  \left({\bf S}\times\frac{\tilde {\bf v}}{\tilde v^0}\right).
\eeq

Notice that the 1PN ${\cal O}(G)$ corrections in (\ref{tildv}, \ref{tild0}) induce ${\cal O}(G^2)$ terms at 2.5PN in Eq. (\ref{spind}) after inserting the LO part of ${\bf A}$ (that is, say for particle one, ${\bf A}^{\rm LO}_1 = {\bf a}^{\rm LO}_1\equiv -G m_2{\bf r}/r^3)$, similarly to what happened in the spin-spin sector \cite{comment,nrgrss,nrgrs2}. 

\section{The next-to-leading order spin-orbit potential}

To compute the spin-orbit potential we proceed systematically. First we compute diagrams with a single graviton exchanged, as in Figs. \ref{onegSO}$a$-$f$. Then we move to the non-linear gravitational effects that can be split in two groups: the {\it seagull} diagrams depicted in Figs. \ref{segSO}$a$-$d$ that follow from Eqs. (\ref{sgh2v4}, \ref{sgh2v5}) and (\ref{h2m}, \ref{h2mv}), and the `three-graviton' interactions shown in Figs. \ref{benzSO}$a$-$g$, where we contract the worldline couplings linear in $H_{\mu\nu}$ with the ${\cal O}(H^3)$ (bulk) interaction from Einstein's action (in background harmonic gauge) \cite{nrgr}. Everywhere in this paper $\int_{\bf p}$ stands for $\int {d^3{\bf p} \over (2\pi)^3}$, and $\int dt$'s are suppressed. The results for the diagrams are collected below.

\subsection{One-graviton exchange}

\begin{figure}[h!]
    \centering
    \includegraphics[width=10cm]{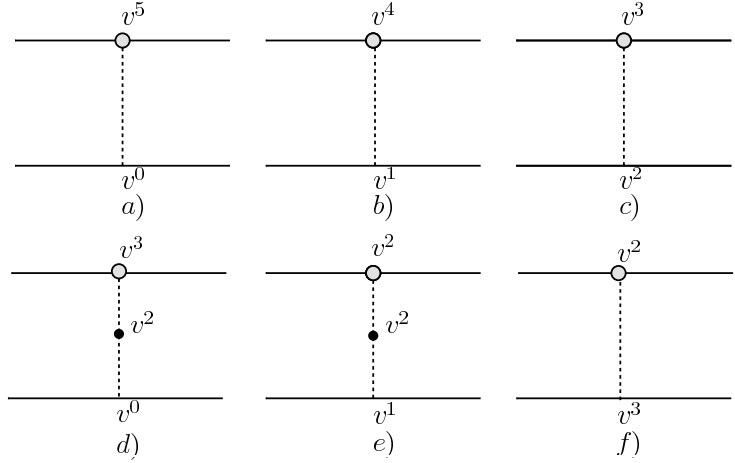}
\caption[1]{\sl One-graviton contributions linear in spin at ${\cal O}(v^5)$. A dashed line represents a potential graviton, a blob account for spin insertions at each PN order from Eqs. (\ref{sgnr1}--\ref{sgnr2}), and a black dot stands for `propagator' corrections as explained in the text.}\label{onegSO}
\end{figure}

It is sometimes convenient to group terms together while computing the one-graviton exchange. For instance, we can obtain most contributions in Figs. \ref{onegSO}$a$-$c$ by (Wick) contracting the ${\cal O}(H)$ spin coupling \cite{nrgrs}
\beq
\label{hsab}
\frac{1}{2 m_p} H_{\mu\nu,\rho}S^{\nu\rho} v^\mu,
\eeq
with the spin-independent part
\beq
-\frac{m}{2m_p} H_{\alpha\beta} v^\alpha v^\beta,
\eeq
and using the two point function \cite{nrgr} 
\beq
\label{prop}
\langle H_{\mu\nu} (x^0_1,{\bf x}_1)H_{\alpha\beta}(x^0_2,{\bf x}_2)\rangle = -i  \delta(x^0_1-x^0_2)
 P_{\alpha\beta\mu\nu} \int_{\bf p} \frac{1}{{\bf p}^2} e^{i{\bf p}\cdot {\bf r}}\eeq
where \beq 
P_{abcd} = \frac{1}{2} \left(\eta_{ac}\eta_{bd}+\eta_{ad}\eta_{bc}-\eta_{ab}\eta_{cd} \right).
\eeq 
The remaining piece comes from a contraction between the terms in Eq. (\ref{sgnr15}) and the one proportional to $H_{00}{\bf v}^2$ in Eq. (\ref{lv2m}). Hence the result for these diagrams is
\beq
V_{\ref{onegSO}abc}= \frac{Gm_2}{r^3}\left[{\bf r}^i\left\{S^{i0}_1\left(\frac{3}{2}{\bf v}_2^2-2 {\bf v}_1\cdot {\bf v}_2\right) + 2 S_1^{ij}{\bf v}_2^j {\bf v}_1\cdot {\bf v}_2 -\frac{{\bf v}_2^2}{2} S_1^{ij}{\bf v}_1^j\right\}+ S^{i0}_1\left(({\bf v}_1-2{\bf v}_2)^i {\bf v}_2\cdot {\bf r}-2{\bf a}_2^i r^2\right)\right], 
\eeq
while the acceleration-dependent piece results from the (partial) time derivative ($\rho =0$) in Eq. (\ref{hsab}). The latter induces derivatives of the particle's position and velocities, as well as its spin. In order to avoid contributions that depend on $\dot S^{i0}_1$, we evaluate them as explained in \cite{nrgrss} (see Eq. (46) in \cite{nrgrss}), namely by using integration by parts so that we may set $\partial_0^{(2)} S^{ab}_1(x^0_1) \equiv 0$.\\

The diagrams in Figs. \ref{onegSO}$de$ account for `propagator' corrections to the LO potential. These arise after we include the non-instantenous part of the propagator in Eq. (\ref{prop}), and expand in powers of $p_0$ \cite{nrgr,eih}
\beq
\frac{1}{p_0^2-{\bf p}^2} = -\frac{1}{{\bf p}^2}\left(1+\frac{p_0^2}{{\bf p}^2}+\ldots \right).
\eeq

Again we face temporal derivatives that we handle as we sketched above (that is the reason we generate ${\bf a}_2$ and $\dot {\bf a}_2$ terms, see below.) Then we have
\bea
V_{\ref{onegSO}de}&=& \frac{Gm_2}{r^3} \left[\left\{S^{i0}_1\left(\frac{1}{2}{\bf v}_2^2-\frac{3}{2r^2} ({\bf v}_2\cdot {\bf r})^2-\frac{1}{2}{\bf a}_2\cdot{{\bf r}}\right)+\left(\frac{1}{2}{\bf v}_2^2 - \frac{3}{2r^2}({\bf v}_2\cdot {\bf r})^2-\frac{1}{2}{\bf a}_2\cdot{\bf r}\right) S_1^{ij}{\bf v}_1^j + 2S_1^{ij}{\bf a}_2^j{\bf v}_2\cdot{\bf r} \right. \right. \nn \\ & & +  \left. \left. r^2S^{ij}_1\dot {\bf a}_2^j +\left(\frac{3({\bf v}_2\cdot {\bf r})^2}{r^2}-{\bf v}_2^2+{\bf a}_2\cdot{\bf r}\right)S_1^{ij}{\bf v}_2^j \right\} {\bf r}^i+S_1^{ij}{\bf v}_2^i{\bf v}_1^j{\bf v}_2\cdot{\bf r}-r^2S^{ij}_1{\bf a}_2^j{\bf v}_2^i- \frac{r^2}{2} S_1^{ij}{\bf v}_1^i{\bf a}_2^j \right. \nn \\ & & + \left.  S^{i0}_1\left({\bf v}_2^i {\bf v}_2\cdot {\bf r} +\frac{1}{2}{\bf a}_2^ir^2\right)\right]
\eea

Finally there is the diagram in Fig. \ref{onegSO}$f$ which is straightforward to calculate,
\beq
V_{\ref{onegSO}f} =   - {\bf v}_2^2 \frac{Gm_2}{r^3} S_1^{jl} {\bf v}_2^l {\bf r}^j.
\eeq

Therefore the total for the one-graviton diagrams reads
\bea
V_{\rm 1grav}&=& \frac{Gm_2}{r^3}\left[\left\{S^{i0}_1\left(2{\bf v}_2^2-2 {\bf v}_1\cdot {\bf v}_2-\frac{3}{2r^2} ({\bf v}_2\cdot {\bf r})^2-\frac{1}{2} {\bf a}_2\cdot {\bf r}\right) + \left(2 {\bf v}_1\cdot {\bf v}_2+\frac{3({\bf v}_2\cdot {\bf r})^2}{r^2}-2{\bf v}_2^2+{\bf a}_2\cdot{\bf r}\right)S_1^{ij}{\bf v}_2^j \right. \right. \nn \\ & & - \left. \left.\left(\frac{3}{2r^2} ({\bf v}_2\cdot {\bf r})^2 +\frac{1}{2} {\bf a}_2\cdot{\bf r}\right)S_1^{ij}{\bf v}_1^j+ 2S_1^{ij}{\bf a}_2^j{\bf v}_2\cdot{\bf r}+r^2S^{ij}_1\dot {\bf a}_2^j \right\}{\bf r}^i+ S^{i0}_1\left(({\bf v}_1-{\bf v}_2)^i {\bf v}_2\cdot {\bf r}-\frac{3}{2}{\bf a}_2^i r^2\right) \right. \nn \\ & & \left. +S_1^{ij}{\bf v}_2^i{\bf v}_1^j{\bf v}_2\cdot{\bf r}-r^2S^{ij}_1{\bf a}_2^j{\bf v}_2^i-\frac{1}{2}r^2S^{ij}_1{\bf a}_2^j{\bf v}_1^i \right]
\eea

\subsection{Non-linear gravitational interactions}

\subsubsection{Seagulls}

\begin{figure}[t!]
    \centering
    \includegraphics[width=10cm]{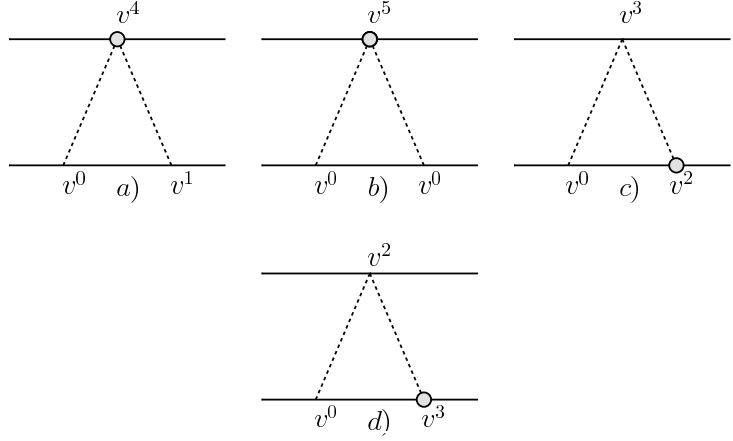}
\caption[1]{\sl Non-linear gravitational effects due the couplings in Eqs. (\ref{sgh2v4}, \ref{sgh2v5}) and (\ref{h2m}, \ref{h2mv}).}\label{segSO}
\end{figure}

These diagrams are shown in Figs. \ref{segSO}$a$-$d$ and account for the ${\cal O}(H^2)$ spin and spin-independent couplings in the worldline action/Routhian. The results are
\bea
V_{\ref{segSO}a}&=& -2\frac{G^2m_2^2}{r^4} S_1^{ij}{\bf r}^j{\bf v}_2^i \\
V_{\ref{segSO}b}&=& 2\frac{G^2m_2^2}{r^4} S_1^{ij}{\bf r}^j{\bf v}_1^i \\
V_{\ref{segSO}c}&=& 2\frac{G^2m_2m_1}{r^4} S_1^{ij}{\bf r}^j{\bf v}_2^i \\
V_{\ref{segSO}d}&=& \frac{G^2m_2m_1}{r^4}\left (S_1^{j0}{\bf r}^j-S_1^{ij}{\bf r}^j{\bf v}_1^i\right) 
\eea
which add up to
\beq
V_{\rm seagulls}= \frac{G^2m_2}{r^4}\left(m_1S_1^{i0}+(m_1-2m_2)S_1^{ij}{\bf v}_1^j+2(m_2-m_1)S_1^{ij}{\bf v}_2^j\right){\bf r}^i 
\eeq

\subsubsection{Three-graviton couplings}

\begin{figure}[t!]
    \centering
    \includegraphics[width=10cm]{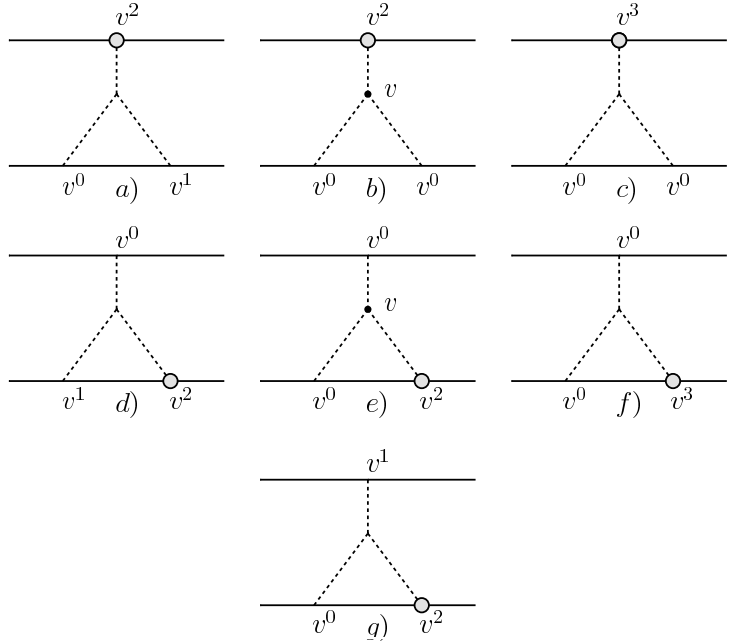}
\caption[1]{\sl Non-linear gravitational effects due to the three-graviton couplings. A black dot represents an extra power of $v$ from a time derivative in the three graviton vertex (see text.)}\label{benzSO}
\end{figure}

Finally we have the three-graviton couplings in Figs. \ref{benzSO}$a$-$g$. These are easily handled with the aid of Mathematica code \cite{nrgrwork}. The results are:

\bea
V_{\ref{benzSO}a}&=&-\frac{G^2m_2^2}{r^4} S_1^{ij}{\bf r}^j{\bf v}_2^i \\
V_{\ref{benzSO}b}&=& \frac{G^2m_2^2}{2r^4} S_1^{ij}{\bf r}^j{\bf v}_2^i \\
V_{\ref{benzSO}c}&=& -\frac{G^2m_2^2}{r^4}\left(2S^{j0}_1{\bf r}^j+ \frac{3}{2}S_1^{ij}{\bf r}^j{\bf v}_1^i\right) \\
V_{\ref{benzSO}d}&=&-3\frac{G^2m_2m_1}{r^4} S_1^{ij}{\bf r}^j{\bf v}_1^i\\
V_{\ref{benzSO}e}&=&\frac{G^2m_2m_1}{2r^4} S_1^{ij}{\bf r}^j{\bf v}_1^i \\
V_{\ref{benzSO}f}&=& \frac{G^2m_2m_1}{r^4}\left(-2S^{j0}_1{\bf r}^j+ \frac{5}{2}S_1^{ij}{\bf r}^j{\bf v}_1^i\right) \\
V_{\ref{benzSO}g}&=&-2\frac{G^2m_2m_1}{r^4} S_1^{ij}{\bf r}^j{\bf v}_2^i, 
\eea
for a total 
\beq
V_{\rm 3grav}= \frac{G^2m_2}{r^4}\left(-2(m_1+m_2)S_1^{i0}+\frac{3m_2}{2} S_1^{ij}{\bf v}_1^j+
\left(\frac{m_2}{2}+2m_1\right)S_1^{ij}{\bf v}_2^j\right){\bf r}^i. 
\eeq

It is easy to foresee that many three-graviton diagrams entail the following integral
\beq
I_{ijl}({\bf r})=\int_{{\bf q},{\bf p}}\frac{{\bf p}^i{\bf p}^j{\bf q}^l}{{\bf q}^2{\bf p}^2({\bf p}+{\bf q})^2}e^{i{\bf p}\cdot{\bf r}}e^{i{\bf q}\cdot{\bf r}} = \int_{{\bf q},{\bf k}}\frac{{\bf p}^i{\bf p}^j({\bf k}^l+{\bf p}^l)}{({\bf p}+{\bf k})^2{\bf p}^2{\bf k}^2}e^{i{\bf k}\cdot{\bf r}} ,
\eeq 
where two powers of momenta come from the three-graviton vertex, and an extra one from the spin coupling(s). To compute $I_{ijl}$ we simply reduce it to scalar integrals, and using
\beq
\int_{\bf q}\frac{1}{{\bf q}^2({\bf p}+{\bf q})^2} = \frac{1}{8|{\bf p}|}, 
\eeq
we obtain
\beq
\label{ijk}
I_{ijl}=\frac{i}{64 \pi^2r^6}\left(r^2({\bf r}^i\delta^{jl}+{\bf r}^j\delta^{il}) - 2 {\bf r}^i{\bf r}^j{\bf r}^l \right).
\eeq
The values for each diagram follow by contracting $I_{ijl}$ with the spin tensor and velocities. Those involving $S^{j0}$ are indeed proportional to (derivatives of) $\langle H_{00}H_{00}H_{00}\rangle$, and can be easily computed from the results in \cite{nrgr}.\\

Aside from the different powers of momenta, the only difference comparing with the kind of diagrams we encountered before in \cite{nrgrss,nrgrs2} for the spin-spin potentials is the appearance of time derivatives in the three-graviton vertex, depicted as a black dot in Figs. \ref{benzSO}$b$ and \ref{benzSO}$e$. For example, we have terms of the sort $\sim H_{00} \partial_i H_{0i} \partial_0 H_{00}$ from Einstein's action which produces a ${\bf p}^i_2 p^0_3$ type of Feynman rule, etc. To handle these diagrams we proceed the same way we compute propagator corrections\footnote{For the three-graviton diagrams in Figs \ref{benzSO}$b$ and \ref{benzSO}$e$ we can ignore the extra terms that are proportional to time derivatives of $S^{ij}$, since these are sub-leading (recall $\dot S^{ij} \sim v^3 S^{ij}$.)}, namely $p^0 \sim {\bf p}\cdot {\bf v}$, which induces an extra power of $v$ in the otherwise (naively) ${\cal O}(v^4)$ diagrams in Figs. \ref{benzSO}$b$ and \ref{benzSO}$e$. 

\subsection{Divergences}

\begin{figure}[h!]
    \centering
    \includegraphics[width=7cm]{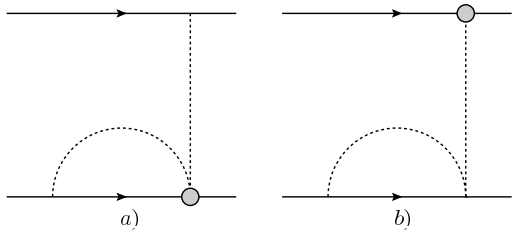}
\caption[1]{\sl Type of divergences in the three-graviton diagrams}\label{div}
\end{figure}

In the computation of the three-graviton contributions we encounter divergences. These can be easily understood diagramatically, as in Figs. \ref{div}$ab$, and arise once a term proportional to ${\bf p}^2_a$ ($a=2,3)$ from the three-graviton Feynman rule cancels out with one of the (two) propagators ($\sim \frac{1}{{\bf p}_a^2}$) that couples the vertex to the worldline. For example we have $I_{iil}=0$ in Eq. (\ref{ijk}) since it produces purely divergent integrals proportional to $\int_{\bf k} {{\bf k}^l \over {\bf k}^2}$ and $\int_{\bf k}{ 1\over {\bf k}^2}$, as shown in Figs. \ref{div}$a$ and \ref{div}$b$ respectively. These divergences are  `self-energy' type of terms that are absorbed into the couplings of the theory, and are set to zero in dimensional regularization (see the discussions in \cite{nrgr,nrgrs,nrgrs2} for more details.)

\subsection{Assembling the pieces}

Adding up all the ingredients the NLO spin-orbit potential reads
\bea
V^{\rm NLO}_{\rm so}&=& \frac{Gm_2}{r^3}\left[\left\{S^{i0}_1\left(2{\bf v}_2^2-2 {\bf v}_1\cdot {\bf v}_2-\frac{3}{2r^2} ({\bf v}_2\cdot {\bf r})^2-\frac{1}{2} {\bf a}_2\cdot {\bf r}\right) + \left(2 {\bf v}_1\cdot {\bf v}_2+\frac{3({\bf v}_2\cdot {\bf r})^2}{r^2}-2{\bf v}_2^2+{\bf a}_2\cdot{\bf r}\right)S_1^{ij}{\bf v}_2^j \right. \right. \nn \\  &-&  \left. \left.\left(\frac{3}{2r^2} ({\bf v}_2\cdot {\bf r})^2 +\frac{1}{2} {\bf a}_2\cdot{\bf r}\right)S_1^{ij}{\bf v}_1^j+ 2S_1^{ij}{\bf a}_2^j{\bf v}_2\cdot{\bf r}+r^2S^{ij}_1\dot {\bf a}_2^j \right\}{\bf r}^i+ S^{i0}_1\left(({\bf v}_1-{\bf v}_2)^i {\bf v}_2\cdot {\bf r}-\frac{3}{2}{\bf a}_2^i r^2\right) \right. \nn  \\ &+& \left. S_1^{ij}{\bf v}_2^i{\bf v}_1^j{\bf v}_2\cdot{\bf r}-r^2S^{ij}_1{\bf a}_2^j{\bf v}_2^i-\frac{1}{2}r^2S^{ij}_1{\bf a}_2^j{\bf v}_1^i \right]\nn \\  &+& \frac{G^2m_2}{r^4} {\bf r}^i\left[-(m_1+2m_2)S_1^{i0}+\left(m_1-\frac{m_2}{2}\right)S_1^{ij}{\bf v}_1^j+\frac{5m_2}{2} S_1^{ij}{\bf v}_2^j\right]
\eea

We can now replace the acceleration terms by using the LO equations of motion, namely ${\bf a}_2={Gm_1\over r^3}{\bf r}$, and including also the LO part of the potential \cite{nrgrs}, we finally obtain
\bea
\label{nloso}
& & V^{\rm so}= \frac{Gm_2}{r^3}\left[\left\{S^{i0}_1\left(1+2{\bf v}_2^2-2 {\bf v}_1\cdot {\bf v}_2-\frac{3}{2r^2} ({\bf v}_2\cdot {\bf r})^2-\frac{G}{r} (3m_1+2m_2)\right)+\right. \right. \nn \\ & & \left. \left. \left(1-\frac{3}{2r^2} ({\bf v}_2\cdot {\bf r})^2+\frac{G}{2r}\left(4m_1-m_2\right)\right) S_1^{ij}{\bf v}_1^j-\left(2-2 {\bf v}_1\cdot {\bf v}_2-\frac{3({\bf v}_2\cdot {\bf r})^2}{r^2}+2{\bf v}_2^2-\frac{G}{2r}\left(2m_1+5m_2\right)\right)S_1^{ij}{\bf v}_2^j \right\}{\bf r}^i \right. \nn \\ & & + \left.  S^{i0}_1({\bf v}_1-{\bf v}_2)^i {\bf v}_2\cdot {\bf r}+S_1^{ij}{\bf v}_1^j{\bf v}_2^i{\bf v}_2\cdot{\bf r}\right] + 1 \leftrightarrow 2, \eea
which can be written as (see Eq. (\ref{split}))
\beq
\label{nloso2}
V^{\rm so}= \sum_{q=1,2} {\bf A}_q \cdot {\bf S}^{(0)}_q + \omega_{{\bf S}_q}\cdot {\bf S}_q 
\eeq
with (for particle 1)
\bea
{\bf A}_1 &=& \frac{Gm_2}{r^3} \left[\left(-1-2{\bf v}_2^2+2 {\bf v}_1\cdot {\bf v}_2+\frac{3}{2r^2} ({\bf v}_2\cdot {\bf r})^2 + \frac{G}{r} (3m_1+2m_2)\right){\bf r} - ({\bf v}_1-{\bf v}_2) {\bf v}_2\cdot {\bf r}\right] \\
\omega_{{\bf S}_1} &=& \frac{Gm_2}{r^3}\left[ - ({\bf v}_2\cdot {\bf r}){\bf v}_1\times{\bf v}_2+\left(1-\frac{3}{2r^2} ({\bf v}_2\cdot {\bf r})^2+\frac{G}{2r}\left(4m_1-m_2\right)\right){\bf L}_1\right. \nn  \\  & & \left. + \left(-2+2 {\bf v}_1\cdot {\bf v}_2+\frac{3({\bf v}_2\cdot {\bf r})^2}{r^2}-2{\bf v}_2^2+\frac{G}{2r}\left(2m_1+5m_2\right)\right){\bf L}_2 \right],
\eea 
and ${\bf L}_{1(2)} = {\bf r}\times {\bf v}_{1(2)}$. 

\section{Spin-orbit dynamics}

Using the potential in Eq. (\ref{nloso}) together with Eqs. (\ref{eomV}, \ref{alssc}, \ref{s0cov}), it is straightforward to obtain the spin dynamics, that is given by the expression (see Eq. (\ref{spind}))
\beq 
\frac{d{\bf S}_1}{dt} = {\bf \omega}_{{\bf S}_1}\times {\bf S}_1 + \left(\frac{\tilde {\bf v}_1}{\tilde v_1^0}\times {\bf S}_1\right)\times {\bf A}_1.
\eeq

\subsection{Precession equation}

To transform the above equation into precession form first we notice that
\beq
{\bf A}_1= \bar {\bf a}_1 + {\bf v}_1\times {\omega^{\rm LO}_{\rm so}}+\frac{1}{2} {\bf v}_1^2 {\bf a}^1_0+\frac{1}{2} {\bf v}_1 ({\bf v}_1\cdot {\bf a}^1_0) \label{a1},
\eeq
with ($\bar {\bf v}_1 \equiv {\tilde {\bf v}_1}/{\tilde v_1^0}$) 
\bea
\label{ta1pn}
\bar {\bf a}_1 \equiv \frac{d}{dt} \bar {\bf v}_1 &=&{\bf a}^1_0 + 
\frac{Gm_2}{r^3}\left[\left(\frac{3}{2r^2}\left({\bf v}_2\cdot{\bf r}\right)^2 -{\bf v}_1^2+4{\bf v}_1\cdot{\bf v}_2-2{\bf v}_2^2+
\frac{G}{r}(2m_2+3m_1)\right){\bf r}\right. \nn \\ &+& ({\bf v}_1-{\bf v}_2)(2{\bf v}_1\cdot{\bf r}-{\bf v}_2\cdot{\bf r})
\left.\frac{}{}\right], 
\eea
${\bf a}^1_0 \equiv -\frac{Gm_2}{r^3} {\bf r}$, and we used the 1PN acceleration from the Einstein-Infeld-Hoffmann Lagrangian \cite{nrgr} plus Eqs. (\ref{tildv}, \ref{tild0}) together with the LO precession frequency
\beq
\label{wsolo} \omega^{\rm LO}_{\rm so} = \omega^{\rm LO}_{\bf S_1} + \frac{1}{2} {\bf v}_1\times {\bf a}^1_0 =  \frac{Gm_2}{r^3}\left( \frac{3}{2} {\bf L}_1 - 2 {\bf L}_2\right). 
\eeq
We may construct now a spin vector with conserved norm with the transformation (see appendix A)
\beq
\label{pnshift}
\bar {\bf S}_1 = \left(1 - \frac{\bar {\bf v}_1^2}{2} -\frac{\bar {\bf v}_1^4}{8}\right){\bf S}_1 +\frac{1}{2} \bar {\bf v}_1 (\bar {\bf v}_1\cdot {\bf S}_1) \left(1+\frac{1}{4}\bar {\bf v}_1^2\right),
\eeq
expanded to NLO. The precession equation thus becomes (up to 2.5PN)\footnote{Notice that the manipulations and results for the spin-orbit frequency are remarkably similar to the results in the spin-spin sector \cite{nrgrss,nrgrs2}.} 
\beq
\label{spindso}
\frac{d}{dt} \bar {\bf S}_1 = \omega^{\rm so}_1\times \bar {\bf S}_1
\eeq
where
\beq
\omega^{\rm so}_1 = \omega_{{\bf S}_1} + \frac{1}{2} \left(1+\frac{{\bf v}_1^2}{4}\right) \bar {\bf v}_1 \times {\bf A}_1 = \omega^{\rm LO}_{\rm so} +\omega^{\rm NLO}_{{\bf S}_1}+\frac{{\bf v}_1^2}{8} {\bf v}_1 \times {\bf a}^1_0 +  \frac{1}{2} {\bf v}_1 \times \bar {\bf A}_1+\frac{1}{2} \left(\bar {\bf v}_1-{\bf v}_1\right) \times {\bf a}^1_0 \label{wson},
\eeq
with $\bar {\bf A}_1 \equiv {\bf A}_1 - {\bf a}_0^1$.\\ 

Hence the spin precession frequency takes the final form
\bea
\label{wso}
\omega^{\rm so}_1 &=& \frac{Gm_2}{r^3}\left[ \left(\frac{3}{2}+\frac{1}{8} {\bf v}_1^2+{\bf v}_2^2-{\bf v}_1\cdot{\bf v}_2-\frac{9}{4r^2} ({\bf v}_2\cdot {\bf r})^2+\frac{G}{2r}\left(m_1-m_2\right)\right){\bf L}_1- \frac{1}{2}({\bf v}_2\cdot {\bf r}){\bf v}_1\times{\bf v}_2  \right. \nn  \\  & & +\left. \left(-2+2 {\bf v}_1\cdot {\bf v}_2+\frac{3({\bf v}_2\cdot {\bf r})^2}{r^2}-2{\bf v}_2^2+\frac{G}{2r}\left(2m_1+3m_2\right)\right){\bf L}_2\right] .
\eea 

\subsection{Comparison with other methods}

The NLO spin-orbit contributions to the dynamics of compact binaries was previously obtained in \cite{bbf}, with the precession frequency given by
\bea
\omega_{\rm BBF}^{\rm so} &=& \frac{Gm_2}{r^3}\left[ \left(\frac{3}{2}+\frac{1}{8} {\bf v}_1^2+{\bf v}_2^2-{\bf v}_1\cdot{\bf v}_2-\frac{9}{4r^2} ({\bf v}_2\cdot {\bf r})^2+\frac{G}{2r}\left(7m_1-m_2\right)\right){\bf L}_1 \right.  \\  & &- \left(\frac{7}{2} ({\bf r}\cdot{\bf v}_2)- 3({\bf v}_1\cdot {\bf r})\right){\bf v}_1\times{\bf v}_2 +\left. \left(-2+2 {\bf v}_1\cdot {\bf v}_2+\frac{3({\bf v}_2\cdot {\bf r})^2}{r^2}-2{\bf v}_2^2+\frac{G}{2r}\left(2m_1+9m_2\right)\right){\bf L}_2\right] \nn . 
\eea 

Notice that it remarkably resembles our result in Eq. (\ref{wso}), but not quite. The frequencies differ by
\beq
\Delta \omega \equiv \omega_{\rm BBF}^{\rm so}-\omega^{\rm so}= \frac{G m_2}{r^3} \left[ \frac{G}{r} 3 m_1 {\bf L}_1+3 m_2\frac{G}{r} {\bf L}_2+3 \left({\bf r}\cdot{\bf v}_1- {\bf v}_2\cdot {\bf r}\right){\bf v}_1\times {\bf v}_2\right],
\eeq
which is a total time derivative
\beq
\label{dw}
\Delta \omega = - 3 \frac{d}{dt} \left(\frac{G m_2}{r} {\bf v}_1\times {\bf v}_2\right).
\eeq

This determines the equivalence of both results after the spin transformation\beq \label{sbbf}
{\bf S}_1 \to {\bf S}_1 - 3\frac{G m_2}{r} ( {\bf v}_1\times {\bf v}_2)\times {\bf S}_1.
\eeq

Moreover, we can use Eq. (\ref{wso}) to construct a Hamiltonian, given by
\beq
\label{hso}
H_{\rm so}({\bf S}_q) = \sum_{q=1,2} \omega^{\rm so}_q\cdot {\bf S}_q ,
\eeq
so that Eqs. (\ref{dw}, \ref{sbbf}) imply that the equations of motion in \cite{bbf} and the ones that derive from the potential in Eq. (\ref{nloso}) fully agree \cite{damour}.

\section{The Newton-Wigner SSC}

Let us consider now the application of the NW SSC at the level of the action. The constraint reads
\beq
S^{i0}_1 = \frac{{\tilde p}^0_1}{{\tilde p}^0_1+m} S^{ij}_1  \frac{{\tilde {\bf p}}^j_1}{\tilde p^0_1}. 
\eeq

Notice this expression is reparameterization invariant, so we have
\beq
\label{pu}
p^\mu = \frac{m}{\sqrt{u^2}} \frac{dx^\mu}{d\lambda} + \cdots, \eeq which we can replace back into the SSC
\beq
S^{i0}_1 = \frac{{\tilde v}^0_1}{{\tilde v}^0_1+\sqrt{{\tilde v}^a_1 {\tilde v}_{1a}}} S^{ij}_1  \frac{{\tilde {\bf v}}^j_1}{\tilde v^0_1}.
\eeq

Then choosing once again $\lambda =x^0$ we obtain (recall ${\bf S}^{(0)i} = S^{0i}$)
\beq
\label{nwssc}
{\bf S}^{(0)}_{1(\rm NW)} = \frac{1}{1+\sqrt{1-\left(\frac{{\tilde {\bf v}}_1}{\tilde v^0_1}\right)^2}}\left({\bf S}_{1(\rm NW)}\times \frac{{\tilde {\bf v}}_1}{\tilde v^0_1}\right).
\eeq

We can now apply Eq. (\ref{nwssc}) into Eq. (\ref{nloso2}) and obtain (suppressing the `NW' in the spin vector)
\beq
\label{nwpot}
V_{\rm NW}^{\rm so}= \sum_{q=1,2} \left\{ \frac{1}{1+\sqrt{1-\left(\frac{{\tilde {\bf v}}_q}{\tilde v^0_q}\right)^2}}
\left(\frac{{\tilde {\bf v}}_q}{\tilde v^0_q}\times {\bf A}_q \right) + \omega_{{\bf S}_q}\right\}\cdot {\bf S}_q, 
\eeq
from which using canonical (Dirac) brackets, that is (say for particle one) $\omega^{\rm so}_{1\rm NW} = \frac{\partial V_{\rm NW}^{\rm so}}{\partial {\bf S}_1}$,
we get the frequency given by Eq. (\ref{wson}) (recall $\bar {\bf v}_1 \equiv \tilde {\bf v}_1/\tilde v_1^0$), and ultimately Eq. (\ref{wso}) including the one `extra term' from the $1/\left(1+\sqrt{1-{\bar {\bf v}}_1^2}\right)$ factor in Eq. (\ref{nwssc}) given by
\beq
\label{extra}
\frac{1}{8} {\bf v}_1^2 \left({\bf v}_1\times {\bf a}^1_0\right),
\eeq
after expanded to ${\cal O}(v^2)$.\\

Let us make a pause and remind ourselves of the form of the transformation between covariant and NW SSCs in Minkowski space given by  Eq. (B13) of \cite{regge}
\beq
\label{b13}
{\bf S}_{\rm NW} = \frac{m}{H_0} {\bf S} +   \frac{{\bf p}\cdot {\bf S}}{H_0(H_0+m)} {\bf p},
\eeq
with $H_0= \sqrt{{\bf p}^2+m^2}$. Using Eq. (\ref{pu}) we have $(\lambda=t)$
\beq
{\bf p} = m \left(1+\frac{{\bf v}^2}{2}\right) {\bf v} + \ldots, ~{\rm and}~~{\bf p}^2 = m^2 \left( {\bf v}^2 + {\bf v}^4+\ldots \right), 
\eeq
 that we can insert back into Eq. (\ref{b13}) and we get
 \beq
 \label{b14}
 {\bf S}_{\rm NW} = \left(1 - \frac{{\bf v}^2}{2}-\frac{1}{8}{\bf v}^4 +\ldots \right){\bf S} +\frac{1}{2} {\bf v} ({\bf v}\cdot {\bf S})\left(1 + \frac{{\bf v}^2}{4} +\ldots \right),
\eeq
which is precisely the transformation in Eq. (\ref{pnshift}) after the replacement ${\bf v} \to {\bar {\bf v}}$. On behalf of the equivalence principle, it is indeed not surprising that the same redefinition as in Minkowski space still applies if rewritten in a locally flat frame and with derivatives taken with respect to proper time (as long as we ignore finite size effects.)

Our result thus provides (yet another, see \cite{proc}) explicit example where the NW SSC leads to standard brackets, this time at linear order on the spin. Hence the Hamiltonian in Eq. (\ref{hso}) can be equivalently obtained by applying the NW SSC from Eq. (\ref{nwssc}) directly in the expression for the potential in Eq. (\ref{nloso2}), such that (see Eq. (\ref{nwpot}))
\beq
H_{\rm so} = \sum_{q=1,2} \left(\omega_{{\bf S}_q} + \frac{1}{1+\sqrt{1-{\bar {\bf v}}_q^2}}\left({\bar {\bf v}}_q\times {\bf A}_q\right)\right)\cdot {\bf S}_q
\eeq

\section{Conclusions}

Using the EFT formalism of NRGR \cite{nrgr} for spinning extended objects \cite{nrgrs,proc}, we computed the NLO contributions to the spin-orbit gravitational potential of inspiralling compact binaries. We explicitly proved the equivalence with previous results in the literature \cite{bbf,damour} and provided extra support for the canonical aspects of the NW SSC (that was already shown to be the case for the spin(1)spin(2) sector to 3PN order in \cite{proc}.) This is consistent with the claims in \cite{brb} and here we extended it to the case of self-gravitating objects, as long as we ignore finite size effects. Moreover, we showed that we can adapt the transformation between NW and covariant SSCs found in \cite{regge} for a spinning particle in a flat background to the (gravitationally bound) binary system.

In a separate paper the results reported here and in \cite{eih,comment,nrgrss,nrgrs2,rad1,spinrad} will be used to compute the spin-orbit and spin-spin contributions to the energy flux and phase evolution to 3PN order \cite{spinrad2}.

\begin{center}
{\bf Acknowledgments}
\end{center}

I would like to thank Delphine Perrodin and Andreas Ross for discussions and checks on Feynman diagrams and potential. This work was supported in part by NSF under Grant No.04-56556.

\appendix
\section{Spin redefinition into precession form}

In this appendix we sketch the steps towards the precession equation.  For the sake of notation, below it should be understood ${\bf v} \to {\bar {\bf v}}_1 \equiv {\tilde {\bf v}}_1/{\tilde v}^0_1 = {\bf v}_1 + \ldots $, and ${\bf a} \to {\dot {\bar {\bf v}}}_1 = {\bf a}^1_{0} + \ldots$, 
see Eqs. (\ref{tildv}, \ref{tild0}, \ref{ta1pn}).

Moreover, all the equations will refer to the dynamics of particle one, even though ${\bf v}_2$ enters in the acceleration as in Eq. (\ref{tildv}). In the spirit of abbreviation let us start by rewriting Eq. (\ref{a1}) as \beq \label{apA} {\bf A} = {\bf a} + {\bf v}\times \omega+ \delta {\bf A},\eeq where $\omega\equiv \omega_{\rm so}^{\rm LO} = \omega_0+\frac{1}{2}{\bf v}\times{\bf a}_{0}$ at LO, with $\omega_0 \equiv \omega_{\bf S}^{\rm LO}$ (see Eq. \ref{wsolo})) and
\beq
\label{remn}
\delta A = \frac{1}{2} {\bf v}^2{\bf a}_0+\frac{1}{2} {\bf v} ({\bf v}\cdot {\bf a}_0).  
\eeq
We will return to this last piece at the end. Let us start with the redefinition (same as in \cite{nrgrs,nrgrss,nrgrs2})
\beq
\label{redef}
\bar {\bf S} = \left(1 - \frac{{\bf v}^2}{2}\right){\bf S} +\frac{1}{2} {\bf v} ({\bf v}\cdot {\bf S}),
\eeq
and move slowly. Taking time derivatives of this expression we have
\beq
{\dot {\bar {\bf S}}} = - ({\bf v}\cdot {\bf a}) {\bf S} +\left(1 - \frac{{\bf v}^2}{2}\right){\dot {\bf S}} +\frac{1}{2} {\bf a} ({\bf v}\cdot {\bf S})+\frac{1}{2}{\bf v} ({\bf a}\cdot {\bf S})+\frac{1}{2} {\bf v} ({\bf v}\cdot \dot {\bf S}),
\eeq
that we can write in terms of ${\bf \bar S}$ after inverting the spin redefinition in Eq. (\ref{redef})
\bea
\dot {\bar {\bf S}} &=& -\frac{1}{2}({\bf v}\times {\bar {\bf S}})\times{\bf a}-\frac{1}{2}({\bf a}\times {\bar {\bf S}})\times{\bf v} +\omega_0\times \left({\bar {\bf S}}-\frac{1}{2}{\bf v}({\bar {\bf S}}\cdot{\bf v})\right)+({\bf v}\times{\bar {\bf S}})\times ({\bf a}+{\bf v}\times {\omega}) \\ & & +\frac{1}{2} {\bf v}~{\bf v}\cdot \left[\omega_0\times {\bar {\bf S}}+ ({\bf v}\times {\bar {\bf S}})\times({\bf a}+{\bf v}\times {\omega})\right]- \frac{{\bf v}^2}{4}\left(({\bf v}\times {\bar{\bf S}})\times{\bf a}+({\bf a}\times {\bar {\bf S}})\times{\bf v}\right)+\frac{1}{4}({\bf a}\times {\bf v})\times{\bf v} ({\bar {\bf S}}\cdot {\bf v}). \nn
\eea

Let us drop the bars from now on and consider only the relevant terms to NLO (recall $\dot {\bf S} \sim v^3 {\bf S}$), then
\bea
\dot {\bf S} &=& \omega_0\times {\bf S}+\frac{1}{2}({\bf v}\times {\bf a})\times{\bf S}-\frac{1}{2}\left(\omega_0+\frac{1}{2}{\bf v}\times {\bf a}_0\right)\times {\bf v}( {\bf S}\cdot{\bf v})-{\bf S}\cdot({\bf v}\times {\omega}){\bf v} \\ & & +\frac{1}{2} {\bf v} \left[\omega_0\times{\bf S}+ ({\bf v}\times{\bf S})\times{\bf a}_0\right]\cdot{\bf v}- \frac{{\bf v}^2}{4}\left(({\bf v}\times{\bf S})\times{\bf a}_0+({\bf a}_0\times{\bf S})\times{\bf v}\right). \nn
\eea

Now we use $[({\bf v}\times {\bf S})\times {\bf a}_0]\cdot {\bf v} = [({\bf v}\times {\bf a}_0)\times {\bf S}]\cdot {\bf v}$, and split it in two halves
\bea
\dot {\bf S} &=& \omega_0\times {\bf S}+\frac{1}{2}({\bf v}\times {\bf a})\times{\bf S}-\frac{1}{2}\left(\omega_0+\frac{1}{2}{\bf v}\times {\bf a}_0\right)\times {\bf v}( {\bf S}\cdot{\bf v})-{\bf S}\cdot({\bf v}\times {\omega}){\bf v} \\ & & +\frac{1}{2} {\bf v} \left[\left(\omega_0+\frac{1}{2}({\bf v}\times{\bf a}_0)\right)\times{\bf S} \right]\cdot{\bf v}- \frac{{\bf v}^2}{4}\left(({\bf v}\times{\bf S})\times{\bf a}_0+({\bf a}_0\times{\bf S})\times{\bf v}\right)+\frac{{\bf v}}{4}\left[({\bf v}\times{\bf a}_0)\times{\bf S} \right]\cdot{\bf v} \nn
\eea
next we identify $\omega = \omega_0 + \frac{1}{2} {\bf v} \times {\bf a}_0$, and we obtain
\bea
\dot {\bf S} &=& \omega_0\times {\bf S}+\frac{1}{2}({\bf v}\times {\bf a})\times{\bf S}-\frac{1}{2}\omega\times {\bf v}( {\bf S}\cdot{\bf v})-{\bf S}\cdot({\bf v}\times {\omega}){\bf v} \\ & & +\frac{1}{2} {\bf v} \left(\omega\times{\bf S} \right)\cdot{\bf v}- \frac{{\bf v}^2}{4}\left[({\bf v}\times{\bf S})\times{\bf a}_0+({\bf a}_0\times{\bf S})\times{\bf v}\right]+\frac{{\bf v}}{4}\left[({\bf v}\times{\bf a}_0)\times{\bf S} \right]\cdot{\bf v}. 
\eea

Hence we realize
\bea
\dot {\bf S} &=& \omega_0\times {\bf S}+\frac{1}{2}({\bf v}\times ({\bf a}+{\bf v}\times\omega))\times{\bf S}- \frac{{\bf v}^2}{4}\left[({\bf v}\times{\bf S})\times{\bf a}_0+({\bf a}_0\times{\bf S})\times{\bf v}\right]+\frac{{\bf v}}{4}\left[({\bf v}\times{\bf a}_0)\times{\bf S} \right]\cdot{\bf v}\nn \\ &+& ({\bf v}\times {\bf S})\times\delta {\bf A}, \nn
\eea
including the term in Eq. (\ref{remn}). In order to have spin evolution with constant norm, our remaining task is to handle the last four terms. Using Eq. (\ref{remn}), these are  
\bea
\delta \dot {\bf S} &=& - \frac{{\bf v}^2}{4}\left[({\bf v}\times{\bf S})\times{\bf a}_0+({\bf a}_0\times{\bf S})\times{\bf v}\right]+\frac{{\bf v}}{4}\left[({\bf v}\times{\bf a}_0)\times{\bf S} \right]\cdot{\bf v}+\frac{1}{2} {\bf v}^2 ({\bf v}\times {\bf S})\times {\bf a}_0 +\frac{1}{2} {\bf v}\cdot {\bf a}_0 ({\bf v}\times {\bf S})\times {\bf v}\nn  \\ &=&   \frac{{\bf v}^2}{4}\left[({\bf v}\times {\bf S})\times {\bf a}_0 -({\bf a}_0\times{\bf S})\times{\bf v}\right]+\frac{{\bf v}}{4}\left[({\bf v}\times{\bf a}_0)\times{\bf S} \right]\cdot{\bf v}+\frac{1}{2} {\bf v}\cdot {\bf a}_0 ({\bf v}\times {\bf S})\times {\bf v}, 
\eea
which can be grouped as
\beq
\delta \dot {\bf S} = \frac{{\bf v}^2}{4} ({\bf v}\times{\bf a}_0)\times{\bf S} + \frac{{\bf v}}{4}\left[({\bf v}\times{\bf a}_0)\times{\bf S} \right]\cdot{\bf v}+\frac{1}{2} {\bf v}\cdot {\bf a}_0 ({\bf v}\times {\bf S})\times {\bf v}.
\eeq
Notice that the first term has already a precession form and, indeed, precisely completes the terms in the expression for ${\bf A}$ in Eq. (\ref{apA}) (the piece proportional to ${\bf v}$ in Eq. (\ref{remn}) cancels out in the cross product), so that 
\beq
\label{dotSa}
 \dot {\bf S} = \omega_0\times{\bf S} + \frac{1}{2} ({\bf v}\times{\bf A})\times{\bf S} + \frac{{\bf v}}{4}\left[({\bf v}\times{\bf a}_0)\times{\bf S} \right]\cdot{\bf v}+\frac{1}{2} {\bf v}\cdot {\bf a}_0 ({\bf v}\times {\bf S})\times {\bf v}.
\eeq
In order to remove the left over terms we need to perform a 2PN shift. Garnering all the ingredients it is now straightforward to show that the following complete transformation (where we restore the bars, see Eqs. (\ref{pnshift}, \ref{b14}))
\beq
\bar {\bf S} = \left(1 - \frac{\bar {\bf v}^2}{2}-\frac{1}{8}{\bar {\bf v}}^4 \right){\bf S} +\frac{1}{2}{\bar {\bf v}} ({\bar {\bf v}}\cdot {\bf S})\left(1 + \frac{\bar {\bf v}^2}{4}\right),
\eeq
removes the unwanted pieces in Eq. (\ref{dotSa}) and introduces one extra term given by Eq. (\ref{extra}), as in Eqs. (\ref{spindso}, \ref{wso}).

\end{document}